\documentclass[twocolumn,floats,floatfix,prb,superscriptaddress,showpacs,aps]{revtex4-1}
\usepackage{graphicx}%
\usepackage{dcolumn}%
\usepackage{bm}%
\usepackage{chemist}

\begin{document}

 \preprint{APS/123-QED}

\title{Verwey transition in single magnetite nanoparticles}

\author{Q. Yu}
\affiliation{Laboratoire de Physique et d'Etude des Mat\'eriaux, UMR 8213, ESPCI-ParisTech-CNRS-UPMC, 10 rue Vauquelin, 75231 Paris, France}
\author{A. Mottaghizadeh}
\affiliation{Laboratoire de Physique et d'Etude des Mat\'eriaux, UMR 8213, ESPCI-ParisTech-CNRS-UPMC, 10 rue Vauquelin, 75231 Paris, France}
\author{H. Wang}
\affiliation{Laboratoire de Physique et d'Etude des Mat\'eriaux, UMR 8213, ESPCI-ParisTech-CNRS-UPMC, 10 rue Vauquelin, 75231 Paris, France}
\author{C. Ulysse}
\affiliation{Laboratoire de Photonique et de Nanostructures, CNRS, 91460 Marcoussis, France}
\author{A. Zimmers}
\affiliation{Laboratoire de Physique et d'Etude des Mat\'eriaux, UMR 8213, ESPCI-ParisTech-CNRS-UPMC, 10 rue Vauquelin, 75231 Paris, France}
\author{V. Rebuttini}
\affiliation{Humboldt-Universit\"{a}t zu Berlin, Institut f\"{u}r Chemie, Brook-Taylor-Str. 2, 12489 Berlin, Germany}
\author{N. Pinna}
\affiliation{Humboldt-Universit\"{a}t zu Berlin, Institut f\"{u}r Chemie, Brook-Taylor-Str. 2, 12489 Berlin, Germany}
\author{H. Aubin}
\email{Herve.Aubin@espci.fr} \affiliation{Laboratoire de Physique et d'Etude des Mat\'eriaux, UMR 8213, ESPCI-ParisTech-CNRS-UPMC, 10 rue Vauquelin, 75231 Paris, France}

\date{\today}

\begin{abstract}
We present a tunnel spectroscopy study of the electronic spectrum of single magnetite \chemform{Fe_3O_4} nanoparticles trapped between nanometer-spaced electrodes. The Verwey transition is clearly identified in the current voltage-characteristics where we find that the transition temperature is electric field dependent. The data show the presence of localized states at high energy, $\varepsilon \sim 0.6~eV$, which can be attributed to polaron states. At low energy, the density of states (DOS) is suppressed at the approach of the Verwey transition. Below the Verwey transition, a gap, $\Delta \sim 300~meV$, is observed in the spectrum. In contrast, no gap is observed in the high temperature phase, implying that electronic transport in this phase is possibly due to polaron hopping with activated mobility.
\end{abstract}

\pacs{81.30.Bx, 71.30.+h, 73.21.-b,73.22.-f}

\maketitle

In contradiction to A.H. Wilson classification\cite{Wilson1931} of solids as insulators or metals according to their band filling, many transition metal oxides with partially filled d-bands are insulating\cite{Boer1937}. As suggested by N.F. Mott\cite{Mott1968}, Coulomb interactions and electron-lattice interactions have an essential role in determining the amplitude of the insulating gap.
While most of the binary metal oxides of the type \chemform{TMO_x} are insulating, a few of them, such as \chemform{VO_2}, \chemform{V_2O_3} and \chemform{Fe_3O_4}\cite{Imada1998}, display a \emph{bad metal} phase in close proximity to the insulating phase, thus offering the opportunity to study electron transport in the presence of strong Coulomb and electron-lattice interactions.

In the regime of strong spatial confinement, phenomena such as quantum confinement effects or non-equilibrium transport can be studied. While semiconductors have been extensively studied in this regime, i.e. quantum dots, only a few studies exist for correlated materials. In a recent set of works by D. Natelson group\cite{Lee2008,Fursina2009,Fursina2010,Fursina2010a,Fursina2012} on magnetite \chemform{Fe_3O_4}, an electric-field induced resistance switching was observed, which has been attributed to the electric-field induced collapse of the insulating gap. Thus, studies in the regime of strong spatial confinement may shed light on fundamental electronic processes in those correlated insulators and metals.

We present here a study of magnetite \chemform{Fe_3O_4} nanoparticles. Magnetite is the archetype of a strongly correlated material\cite{Edwards1985,Imada1998,Garcia2004,Walz2002}, it has the cubic reverse spinel \chemform{AB_2O_4} structure, where tetrahedral A sites are occupied by \chemform{Fe^{3+}} cations, octahedral B sites are occupied by cations of valence between \chemform{Fe^{3+}} and \chemform{Fe^{2+}}.


At room temperature, the conductivity of magnetite is weak, $\sigma\sim 50~S~cm^{-1}$, Ref.~\cite{Kuipers1979}, much lower than the Ioffe-Regel\cite{Ioffe1960} minimum of conductivity for a metal, $\sigma_{min}\sim e^2/3\hbar a\sim 3000~S~cm^{-1}$. Electronic transport has been attributed to charge fluctuations between \chemform{Fe^{3+}} and  \chemform{Fe^{2+}} sites\cite{Verwey1947}. Upon decreasing the temperature, a first order phase transition, where the resistivity increases by two orders of magnitude, is observed at the Verwey temperature $T_V\sim120~K$\cite{VERWEY1939}. This transition was initially interpreted by E.J. Verwey as the consequence of the ordering of \chemform{Fe^{3+}} and \chemform{Fe^{2+}} in alternated (001) planes\cite{Verwey1947}.

Cullen and Callen\cite{Cullen1970} suggested that charges are transported by band-like delocalized states for $T>T_V$ and that a collective Coulomb gap opens for $T<T_V$. In contrast, Mott\cite{Mott1980,Austin1969}, Chakraverty\cite{Chakraverty1974} and Yamada\cite{Yamada1980} suggested that polarons, formed because of strong electron-phonon interactions, are responsible for charge transport. In this polaronic model, N.F. Mott\cite{Mott1970}, suggested that the Verwey transition has much similarities with the Wigner crystallization\cite{Wigner1938}. At low temperature, Coulomb interactions localize the polarons, while at high temperature, the polaron crystal melts, allowing charge transport through hopping.

For $T<T_V$, DC conductivity, infra-red optical conductivity\cite{Park1998,Gasparov2000} and terahertz conductivity\cite{Pimenov2005} decrease with temperature. DC transport follows an Arrhenius law with an activation energy about $\sim 120~meV$\cite{Kuipers1979}. In the optical conductivity spectrum, a peak at mid-infrared, $\varepsilon\sim 0.5~eV$, has been interpreted as the possible signature of polarons \cite{Austin1969,Devreese2009,Basov2011}. This spectral feature is also visible in photoemission\cite{Park1997,Schrupp2005} and STM spectroscopy measurements\cite{Poddar2003,Jordan2006}. In this insulating phase, photoemission\cite{Chainani1995,Park1997,Schrupp2005} and tunnel spectroscopy\cite{Jordan2006} indicate the disappearance of the DOS within an energy range $\varepsilon\sim 100-300~meV$ around the Fermi level.

For $T>T_V$, the nature of electronic transport is still poorly understood. DC transport also follows an Arrhenius law with a smaller activation energy about $10-100~meV$\cite{Kuipers1979,Gasparov2000,Pimenov2005}. The activated temperature dependence of infra-red optical conductivity was described within a polaronic model\cite{Gasparov2000}, however, terahertz measurements by A. Pimenov {\it et. al.}\cite{Pimenov2005} suggest the presence of a Drude peak, implying band-like transport. As the conductivity can be generally described as $\sigma=n e \mu$, where $n$ is the electron density and $\mu$ is the electronic mobility, the Arrhenius law could be either due to an activated carrier density, $n=n_0 \exp(-\Delta/k_BT)$, as expected for a semiconductor with a band gap $\Delta$, or an activated mobility, $\mu=\mu_0 \exp(W_P/2k_BT)$, as expected for polarons of binding energy $W_P$ with respect to the Fermi energy.

In this context, detailed studies of the DOS above $T_V$ is required to test the existence of the gap and discriminate between the two models. The observation of a gap in the DOS would indicate that electronic transport is due to delocalized states, the absence of this gap would suggest instead that electronic transport is due to hopping of polaron excitations. While early photoemission measurements\cite{Chainani1995} indicated a finite DOS at the Fermi level, other measurements suggested the presence of a small gap $\sim 50~meV$\cite{Park1997,Schrupp2005}.

With its very high resolution, tunnel spectroscopy is ideally suited for the study of the DOS\cite{Mottaghizadeh2014,Kuemmeth2008,Jordan2006,Poddar2003,Aubin2002}. A first scanning tunnel spectroscopy experiment on magnetite nanoparticles\cite{Poddar2003} led to the observation of a gap $\sim 0.2~eV$ in the DOS for $T<T_V$, which was found to be replaced by a peak structure for $T>T_V$. In another experiment on single crystals\cite{Jordan2006}, a gap $\sim 0.2~eV$ independent of temperature has been observed in the DOS.

In tunneling experiments, because the measured Differential Conductance $dI/dV$ Curve (DCC) depends on the DOS $N(\varepsilon)$ and the transmission coefficient $T(\varepsilon)$ of the tunnel barrier, one major issue is to establish the correct base conductance curve to which the DCC will be normalized. Indeed, this normalization is required to obtain the DOS curves from the DCC curves. This difficulty is alleviated in superconductors\cite{Aubin2002} where sharp features are present in the DOS, however, in complex materials where features of interest in the electronic spectrum are broad and spread on a large energy range from $0$ to $\sim 1~eV$, the interpretation of the DCCs may be impossible. For this reason, measurements of the DCCs as function of temperature is required to study the evolution of the DOS as the Verwey transition is crossed.

To that end, we trapped single magnetite nanoparticles between nanometer-spaced electrodes deposited on a chip circuit. In comparison to scanning tunnel spectroscopy, such junctions are highly stable and can be measured as function of temperature. In comparison to planar junctions, where an inert tunnel barrier is often difficult to fabricate, the presence of small ligands at the surface of chemically synthesized nanoparticles allows for the formation of high impedance tunnel barriers, as required for tunnel spectroscopy. On-chip spectroscopy proved to be a reliable technique as shown by numerous experiments of electronic transport across single molecules\cite{Vincent2012} or the observation of single electronic levels in gold nanoparticles\cite{Kuemmeth2008}. Furthermore, non-equilibrium transport can also be studied as a large electric field can be applied on the nanoparticle.

\begin{figure}[ht!]
  \begin{center}
    \includegraphics[width=7cm]{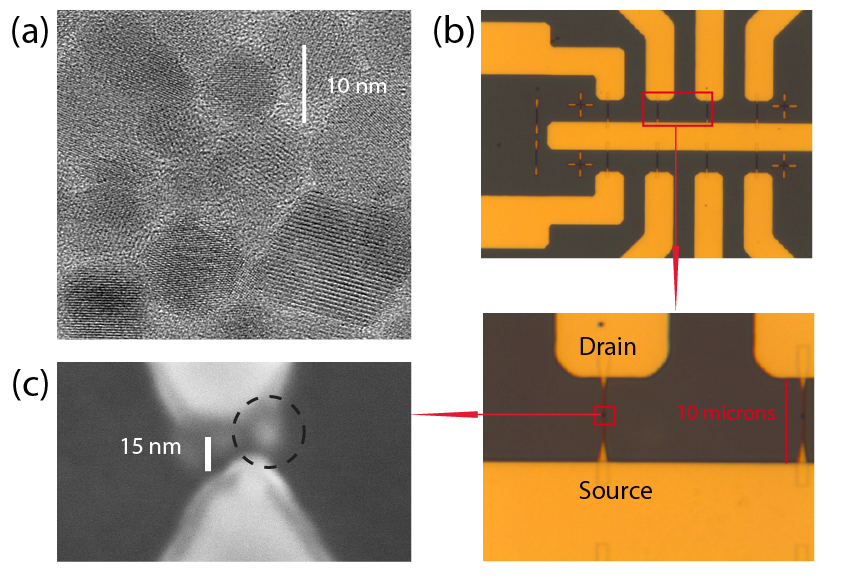}
    \caption{\label{Fig1} a) TEM image of \chemform{Fe_3O_4} nanoparticles. b) Optical microscopy images of circuits. c) High resolution SEM image, $\times 800 000$, of a nanoparticle trapped between two electrodes.}
   \end{center}
\end{figure}

\chemform{Fe_3O_4} nanoparticles are obtained by the reaction of iron (III) acetylacetonate in benzyl alcohol at $175^\circ C$ followed by ligand exchange with citric acid. The citric acid allows the electrostatic stabilization of nanoparticles in water and contributes to the formation of the tunnel barrier\cite{SupplementalMaterial}.  This synthesis provides populations of nanoparticles within the size range $d=6-12~nm$, as shown by the TEM picture, Fig 1a. (Cf. also Figure S1, S2, S3, and discussion in the supporting information). While better nanoparticle size dispersions can be obtained with synthesis in organic solvents, they present the major inconvenient of employing long alkyl chains surfactants for colloidal stability, which produce thick insulating barrier on the nanoparticles and make them inappropriate for the fabrication of junctions.

\begin{figure*}[h!]
  \begin{center}
    \includegraphics[width=18cm]{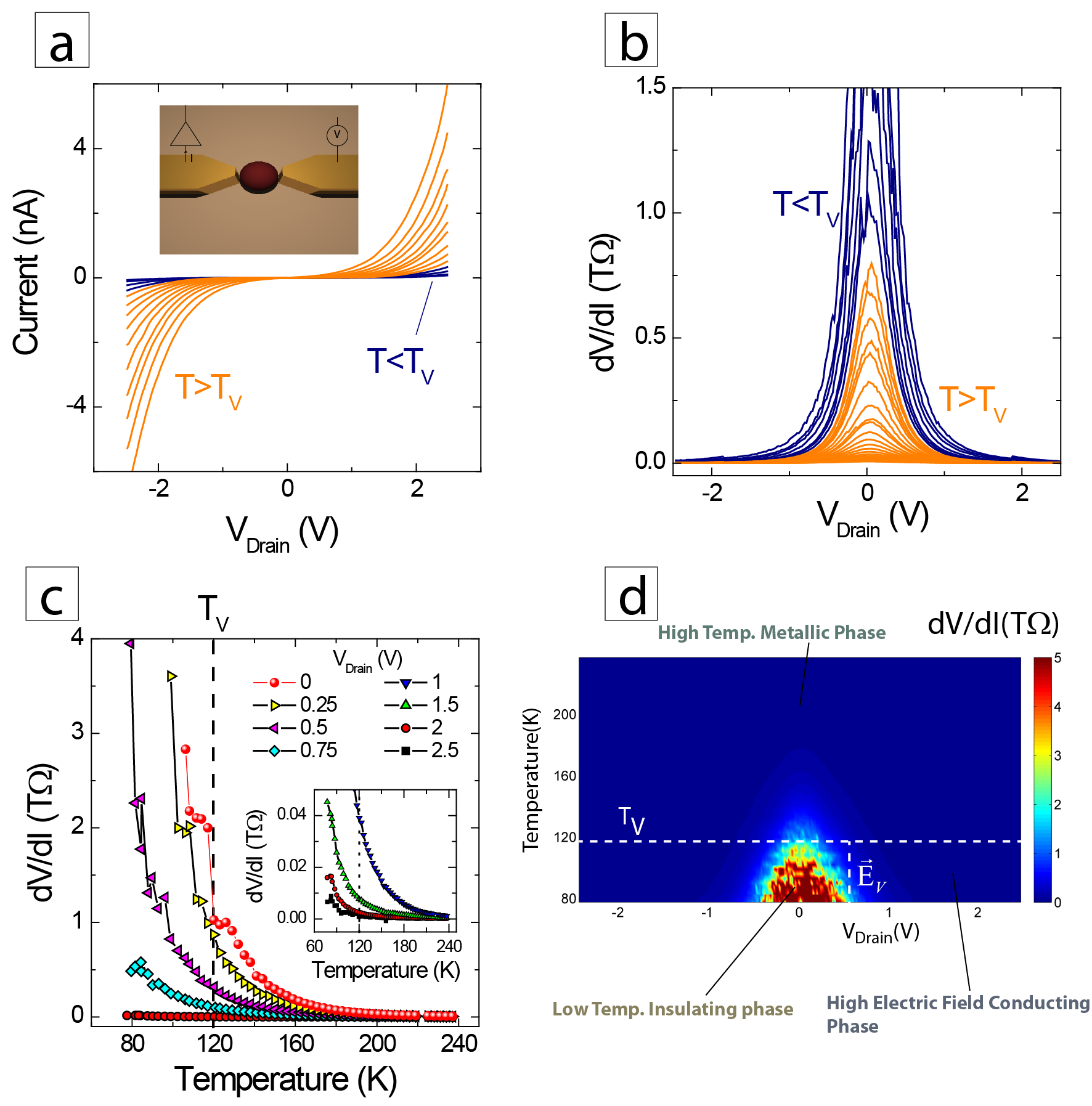}
    \caption{\label{Fig2}
    a) IV curves of a single nanoparticle junction. b) $dV/dI$ as function of voltage. c) $dV/dI$ as function of temperature at selected drain voltages. The dashed line shows the Verwey transition temperature $T_V \sim 120~K$. d) Color plot of $dV/dI$ displaying the out-of-equilibrium phase diagram for the Verwey transition.}
   \end{center}
\end{figure*}

By ebeam lithography, we fabricate chip circuits, each one contains 32 nanogaps, where the drain and source electrodes are separated by a distance of about $10~nm$, Fig 1b. On this chip, we trap a single nanoparticle within a nanogap following a method described previously\cite{Yu2013b}.  In this method, the chip is maintained in high vacuum, $10^{-6}~mbar$, and the nanoparticles are projected through a fast pulsed valve. After each projection, where a small amount of nanoparticles is deposited, the tunnel current is measured to check for the presence of a nanoparticle. The projection is repeated hundred of times until a nanoparticle is detected, as shown Fig. S4. Representatives high resolution SEM images of nanoparticle devices are shown Fig. 1c and Fig. S5. Five chip circuits have been measured from $T=300~K$ to $T=4.2~K$, which have similar IV curves. The projection technique described here has significant advantages. First, because the sample is fabricated in high vacuum and the process is started with insulating junctions, i.e. unbridged nanogaps, there is no doubt that the conducting channels are formed with the magnetite nanoparticles. Second, the setup is implemented in a glove box under argon. Thus, after fabrication, the sample can be kept free from oxidation indefinitely. Indeed, magnetite oxidizes into the insulating hematite, \chemform{Fe_2O_3}, in presence of oxygen. We found that only magnetite nanoparticles can lead to the formation of conducting junctions. No conducting junctions could be obtained with oxidized magnetite solutions. Furthermore, a conducting junction turns into an insulating junction when kept a few hours in oxygen gas.

Fig. 2a shows the IV curves of a typical sample as a function of temperature. Taking the numerical derivative of these IV curves provides the differential resistance $R=dV/dI$ as a function of voltage $V_{Drain}$, at different temperatures between $T\sim75~K$ and $T\sim240~K$, Fig. 2b. From these curves, one can extract $dV/dI$ as function of temperature, at different voltage values between $V_{Drain}=0~V$ and $V_{Drain}=2.5~V$, Fig. 2c. For $V_{Drain}=0~V$, the resistance increases at the Verwey temperature, $T_V\sim 120~K$, which is identical to the transition temperature observed in bulk magnetite. Upon increasing $V_{Drain}$, the transition temperature decreases. A color plot of $dV/dI$ curves as function of temperature and voltage, Fig. 2d, provides a phase diagram for this nanoparticle and demonstrates that the phenomena of charge localization at the Verwey transition also occurs in a single nanoparticle.

In this insulating phase, the Debye screening length is larger than the nanoparticle diameter, $\lambda_D > 1~\mu m$ at $T\sim75~K$, Fig.~S6, implying that a large electric field exists in the nanoparticle. In this situation, the IV curve represents an out-of-equilibrium transport measurement where the Verwey transition is shifted to lower temperature due to out-of-equilibrium effects. Unlike previous measurements on thin films and nanoparticles films, where the contact resistance between the electrodes and the magnetite was low, no {\it switching} is observed in the IV curves\cite{Lee2008,Fursina2009,Fursina2012}. Furthermore, no hysteresis is observed as function of voltage, indicating the absence of heating effects in these junctions where the tunnel current is extremely small. Finally, the IV curves are reproducible as the temperature is cycled between low temperature and room temperature.

In the conducting phase, the Debye screening length is smaller than the nanoparticle diameter, $0.4 ~nm\lesssim \lambda_D \lesssim 1.8 ~nm$, Fig.~S6, implying that the IV curves can be interpreted as tunnel characteristics. A Fowler-Nordheim plot, Fig.~3a, allows identifying the different tunneling regimes.

\begin{figure}[ht!]
  \begin{center}
    \includegraphics[width=8cm]{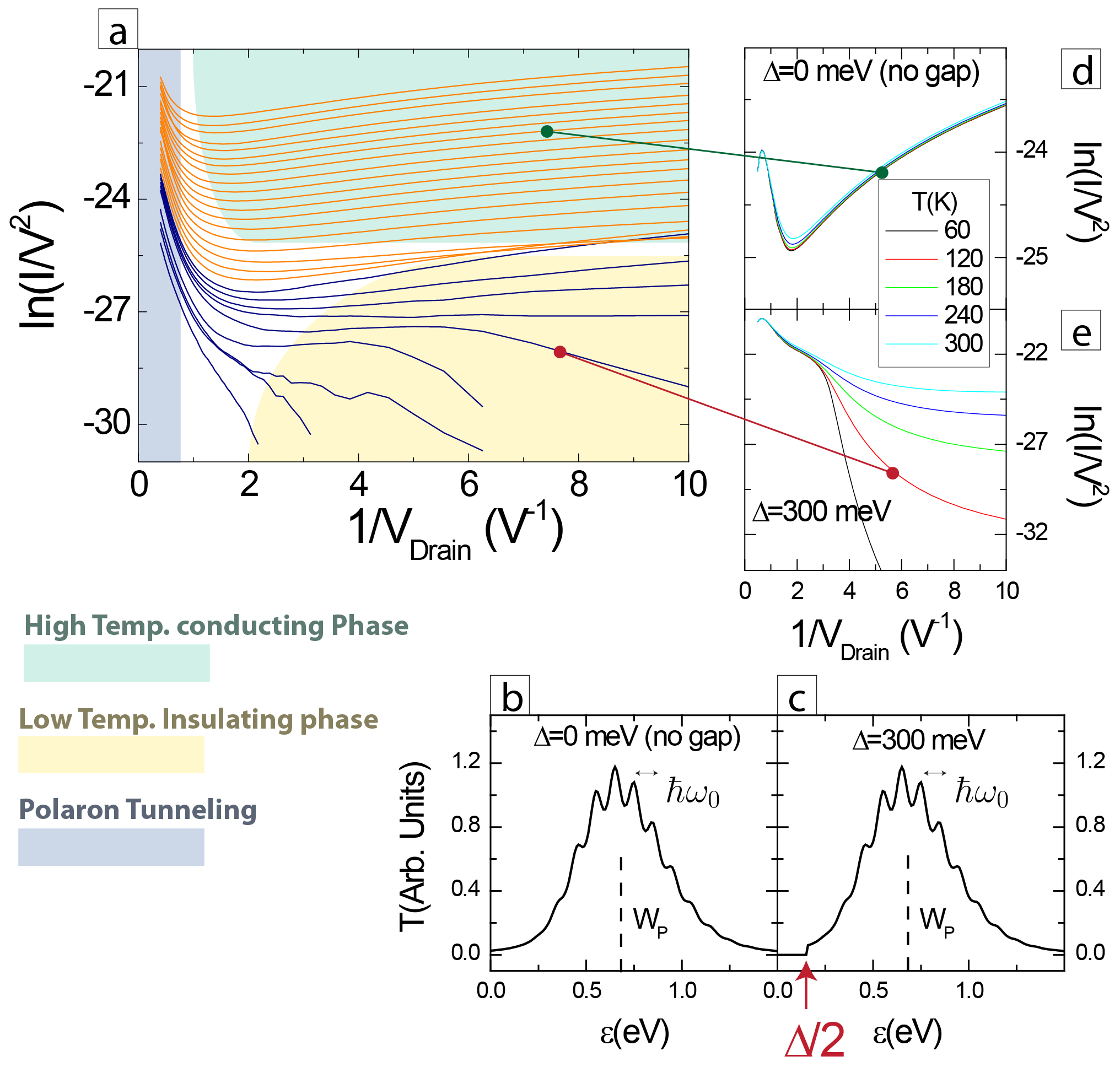}
    \caption{\label{Fig3} a) Fowler-Nordheim plot of the IV curves shown Fig. 2a. Three regimes can be distinguished, see text. b) Transfer function for a tunnel barrier containing a polaron state at the energy $W_P\sim0.6~eV$. The subpeaks are separated by the phonon energy $\hbar \omega_0 \sim 100~meV$. c) Transfer function for a tunnel barrier containing a polaron state at the energy $W_P\sim0.6~eV$ and a gap $\Delta=300~meV$ at the Fermi level. d) Fowler-Nordheim plot of calculated IV curves at temperatures between $T=60~K$ and $T=300~K$ with no gap at the Fermi level. e) Fowler-Nordheim plot of calculated IV curves at temperatures between $T=60~K$ and $T=300~K$ with a gap $\Delta=300~meV$ at the Fermi level.
}
\end{center}
\end{figure}

To summarize figure~3a, three distinct regimes can be identified, which are indicated by distinct colors. In the green zone, at low voltage,  $V_{Drain}<1~V$, and high temperature, $T>T_V$, the curve has a positive slope indicating a linear tunneling regime, $I\sim V$, due to a finite DOS at the Fermi level of the nanoparticle. In the yellow zone, at low voltage, $V_{Drain}<1~V$, and low temperature, $T<T_V$, the slope becomes negative in the insulating phase. In the blue zone, at high voltage, the curve has a negative slope indicating, in analogy with transport measurements in single molecular devices\cite{Beebe2006}, the presence of localized states. In single molecules, the localized states are the HOMO and LUMO levels, in magnetite, they are most likely polaron states of binding energy $W_P\sim 0.6~eV$ which have been identified in photo-emission\cite{Schrupp2005} and optical conductivity spectra\cite{Park1998,Gasparov2000}.

\begin{figure}[ht!]
  \begin{center}
    \includegraphics[width=8cm]{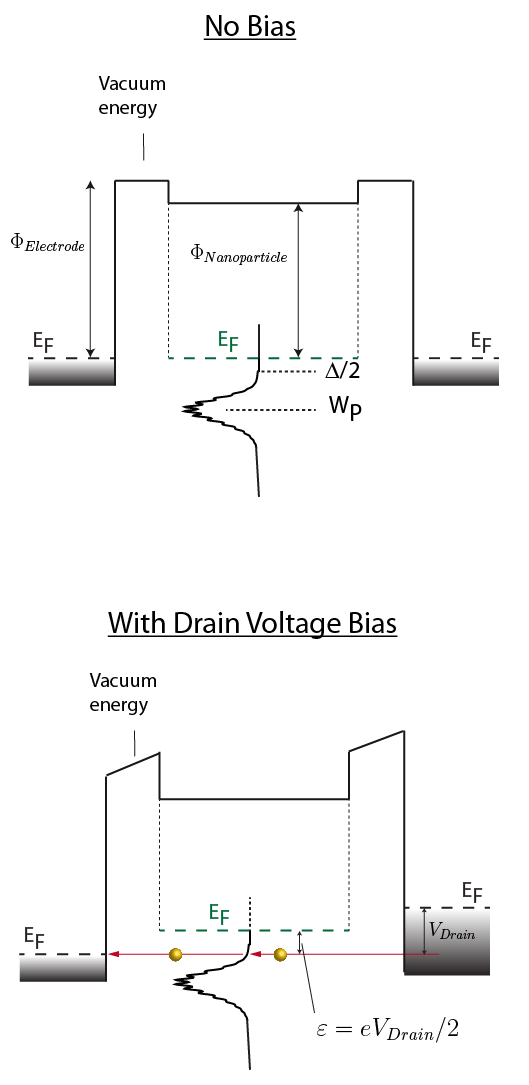}
    \caption{\label{Fig4} Schematic of the junction with no applied bias (left) and applied bias (right). The different energy scales shown are the polaron binding energy W$_P$, the fermi energy E$_F$, the work function of the electrodes $\Phi_{Electrode}$ and the work function of the nanoparticle $\Phi_{Nanoparticle}$.
}
\end{center}
\end{figure}

For a polaron of binding energy $W_P$, resulting from the interaction between an electron and a phonon of energy $\hbar \omega_0$, the transmission coefficient has been determined theoretically\cite{Lundin2002}:

\begin{eqnarray}\label{Eq:polaron}
\nonumber T_p(\varepsilon)&=& e^{-g_1[1+2n_B]}\\
\nonumber&\times& \sum_{\ell=-\infty}^{\infty}I_{\ell}(2g_1\sqrt{n_B[1+n_B]})e^{-\ell\hbar\omega_0\beta/2}\\
&\times&\frac{\Gamma^2}{[\varepsilon+g_1\hbar\omega_0+\ell\hbar\omega_0]^2+\frac{\Gamma^2}{4}}
\end{eqnarray}

where $I_{\ell}$ is a modified Bessel function, $\Gamma$ represents level broadening due to the coupling with the electrodes, $\hbar \omega_0$ is the characteristic energy scale of the phonons, $g_1$ describes the electron-phonon coupling and $n_B=1/(\exp(\beta\hbar\omega_0)-1)$ is the Bose distribution.

Using similar parameters as those used for the analysis of photoemission data\cite{Schrupp2005}, $\hbar \omega_0=0.1~eV$, $g_1=3.5$, and $g_2=\Gamma/\hbar \omega_0=1$, the transmission coefficient displays a broad peak with a maximum at $W_P\sim 0.6~eV$, Fig.~\ref{Fig3}b, c. This broad peak is composed of multiple sub-peaks separated by the phonon energy $\hbar \omega_0$. The sub-peaks are the consequence of the emission and absorption of phonons by the tunneling electron.

As the nanoparticle is connected to the electrodes by two tunnel junctions, the full transmission coefficient of the junction is given by :

\begin{equation}\label{Eq:transmission}
T(\varepsilon)=t_b^2\times T_p(\varepsilon)
\end{equation}

where we describe the transmission coefficient $t_b$ of the tunnel barrier by a small  constant coefficient, $t_b<<1$, independent of applied voltage. The value of this transmission coefficient depends on details of the tunnel barrier such as the distance between the electrodes and the nanoparticle as well as the work functions, $\Phi_{Nanoparticle}$ and $\Phi_{Electrode}$.
The overall magnitude of the tunnel current in a given device can be limited by the tunnel barrier, however, the barriers themselves are not expected to introduce voltage-dependent structures in the IV curves. First, the applied voltage remains small enough with respect to the work functions which are of the order of $\Phi\sim 5~eV$. Second, in the analysis of the DOS curves shown below, the data will be normalized to the high temperature curve to remove the effects of the tunnel barrier, which is a common practice\cite{Aubin2002}.


We can know calculate the I(V) characteristics with the Landauer relation :

\begin{equation}\label{Landauer}
   I(V)=\frac{2e}{h}\int_{-\infty}^\infty T(\varepsilon) \left[f_L(\varepsilon)-f_R(\varepsilon)\right]d\varepsilon
\end{equation}

where $f_L(\varepsilon)=\frac{1}{\exp\left[\frac{\varepsilon-\mu_L-e\phi_L}{k_BT}\right]+1}$ and $f_R(\varepsilon)=\frac{1}{\exp\left[\frac{\varepsilon-\mu_R-e\phi_R}{k_BT}\right]+1}$ are the Fermi level distributions for the left electrode (L) and right electrode (R), respectively, and $V=\phi_L-\phi_R$ is the polarization voltage of the junction which is split equally between the left and right junctions as shown Fig.~4. To describe the amplitude of the conductance of the junction, we need to assume that the transmission coefficient of the tunnel barrier is $t_b\sim 0.0025$. The drain bias $V_{Drain}$ leads to a difference between the energy levels of the nanoparticle and the electrodes given by $\varepsilon=eV_{Drain}/2$. This calculation provides the I(V) characteristics, shown Fig.~3d, which reproduce the sign change of the slope and the position of the minimum in the Fowler-Nordheim plot. We see that these curves do not change with temperature. However, experimentally, upon decreasing the temperature, the slope at low bias changes sign across the Verwey transition, Fig~3a.  This evolution with temperature indicates a modification of the DOS in the nanoparticle as a consequence of charge localization at the Verwey temperature.

\begin{figure*}[ht!]
  \begin{center}
    \includegraphics[width=18cm]{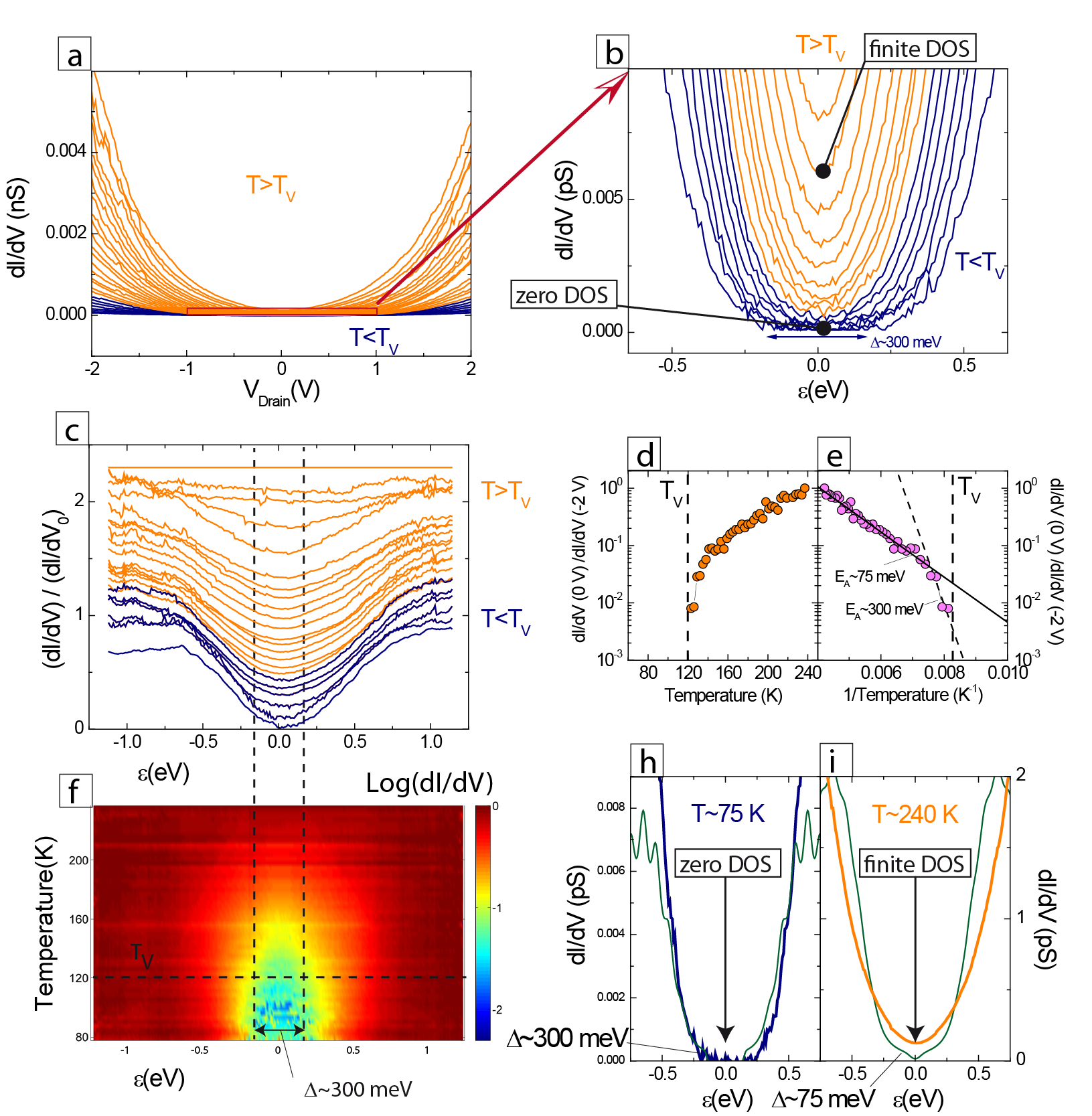}
    \caption{\label{Fig5} a) $dI/dV$ as function of temperature, from $T\sim 240~K$ to $T\sim 75~K$. On the panels b) c) e) and f), the voltage scale has been replaced by the energy scale $\varepsilon =eV_{Drain}/2$. b) Zooming shows that $dI/dV$, at $V_{Drain}=0$, remains finite for $T>T_V$, but decreases quickly to zero for $T<T_V$. From these curves, the gap can be estimated to be about $\Delta\sim 300~meV$ in the insulating phase. c) Normalized conductance $[dI/dV (T)] / [dI/dV(T=240~K)]$. The curves have been shifted for clarity. d) Normalized conductance at $V_{Drain}=0~V$ as function of temperature. e) Arrhenius plot of the Normalized conductance at $V_{Drain}=0~V$. The full circles are the experimental data. The continuous line is a fit that provides the activation energy $E_A\sim 75~meV$. The dashed line represents the expected temperature dependence, at temperatures $T<T_V$, of the zero bias conductance for the activation energy $E_A\sim 300~meV$. This last value for the activation energy is obtained from the energy spectra on panel b). f) Color plot of the normalized conductance shown in c). It shows the opening of the gap $\Delta\sim 300~meV$ at the approach of the Verwey transition. h) $dI/dV$ measured experimentally at $T\sim 75~K$ (blue curve) and calculated (dark green curve) using the polaronic transfer function including a gap $\Delta=300~meV$.
    i) $dI/dV$ measured experimentally at $T\sim 240~K$ (orange curve) and calculated (dark green curve) using the polaronic transfer function including a gap $\Delta=75~meV$.
}
   \end{center}
\end{figure*}

From the relation~\ref{Landauer}, it can be shown that the differential conductance is proportional to the DOS, $dI/dV(V) \propto N(eV)$, in the zero bias limit. Fig.~5a shows the $dI/dV$ curves at different temperatures, from $T\sim 240~K$ to $T\sim 75~K$. On Fig.~5b 5c and 5d the voltage scale has been replaced by the energy scale $\varepsilon =eV_{Drain}/2$. A zoom on $dI/dV$,  at $V_{Drain}=0$, Fig.~5b, shows that the DOS remains finite at any temperature $T>T_V$ but quickly decreases to zero for $T<T_V$. From these curves, we can estimate the magnitude of the gap in the insulating phase, $T<T_V$, to be of the order of $\Delta\sim 300~meV$. The amplitude of which is similar to the one obtained from photoemission measurements\cite{Chainani1995,Park1997,Schrupp2005} infra-red conductivity\cite{Park1998,Gasparov2000} and terahertz conductivity\cite{Pimenov2005}.

To remove the effect of the tunnel barrier in the analysis of $dI/dV$ curves, it is usual to normalize them with a base curve $dI/dV_0$ measured at the highest possible temperature\cite{Aubin2002}, which provides a valid representation of the DOS when the applied bias is small compared to the barrier height.

\begin{equation}\label{Eq:DOS}
\frac{dI/dV_T(V)}{dI/dV_0}=\int{\frac{N(\epsilon)}{N_0}\frac{\partial f(\epsilon-eV,T)}{\partial eV}d\epsilon}\sim \left[\frac{N(\epsilon)}{N_0}, V \rightarrow 0\right]
\end{equation}

Thus, plotting $N(\varepsilon)\sim [dI/dV (T)] / [dI/dV(T=240~K)]$ provides the DOS of the nanoparticle as function of temperature, Fig.~5c. From these DOS curves, one extract the DOS at zero energy as function of temperature, shown Fig.~5d and Fig.~5e. The Arrhenius representation, Fig.~5e, shows that at high temperature, $T>T_V$, the DOS decreases with temperature following an Arrhenius law with the activation energy $E_A\sim 75~meV$. At the approach of the Verwey transition, Fig.~5d shows that the DOS, measured at zero bias, decreases to a very low value, below the noise level of the current to voltage amplifier. In the insulating phase, for $T<T_V$, the color plot, Fig.~5f, shows the opening of the gap $\Delta\sim 300~meV$ that we already identified in Fig.~5b and Fig.~5c.


Knowing the gap value for $T<T_V$, it can be introduced in the transfer function by multiplying Eq.~\ref{Eq:polaron} with a Heaviside function $\mathcal{H}(\varepsilon-\Delta)$ as shown Fig.~3c. Note that the introduction of this gap does not shift the binding energy of the polaron. Fig.~3e shows the resulting IV curves in a Fowler-Nordheim representation. They have a negative slope at any voltage and are temperature dependent as observed experimentally. This calculation can also be compared with the $dI/dV$ curves, Fig.~5h, where one find that this transfer function qualitatively describes the experimental data and the suppression of conductance at zero bias.

For $T>T_V$, if we interpret the activation energy as the consequence of the opening of a gap of value $\Delta\sim 75~meV$, the DOS plotted as function of energy should be suppressed to zero at low energy, $\varepsilon < \Delta$, even at high temperature $T\sim 240~K$, as shown by the theoretical calculation, Fig.~5i, where the gap $\Delta\sim 75~meV$ has been introduced in the transfer function. Thermal broadening alone would not be able to obscure the gap in the DOS until a temperature $T\sim\Delta/k_B\sim 900~K$ is reached. As no gap is observed in the voltage dependence of the $dI/dV$ curves, this implies that no gap larger than $\Delta>10~meV$ exists in the high temperature phase.

Thus, this activated law, as also observed in DC transport of single crystals\cite{Kuipers1979,Pimenov2005} and optical conductivity\cite{Gasparov2000} cannot find its origin in the existence of a gap in the DOS. The most likely explanation is that it originates from an activated mobility  as expected for polaron hopping. Such a scenario was already considered for the interpretation of DC transport\cite{Kuipers1979}, optical conductivity measurements\cite{Gasparov2000} and contact resistance phenomenology\cite{Fursina2010}. This scenario seems to be in contradiction with recent photoemission measurements\cite{Park1997,Schrupp2005}. The elucidation of this contradiction should await the availability of higher resolution photoemission measurements.

To summarize, measurements of the temperature dependence of the differential resistance of a nanogap structure bridged by a single magnetite nanoparticle demonstrate the presence of the Verwey transition in a single magnetite nanoparticle. At low bias, the normalized differential conductance curves can be interpreted as DOS spectra and provide a temperature-voltage map of the DOS. This map suggests the opening of a gap as the temperature is reduced below the Verwey transition, with
a maximum gap value of $\Delta=300~meV$  being reached for temperatures below $T\sim75
~K$. In the conducting phase, $T>T_V$, no gap exists in the DOS despite the observation that the differential conductance at zero bias follows an Arrhenius law with the activation energy $E_A\sim 75~meV$. In a model where electronic transport is due to delocalized states, the observation of an activated law would imply the existence of an band gap with activated carrier density. The observation that no gap is present in the DOS for $T>T_V$ suggests instead that electronic transport is due to polaron hopping with activated mobility. Finally, the signature of the polarons has been identified as the upturn observed in IV curves plotted in the Fowler-Norheim representation.


\bibliography{Bibliography}

\begin{thebibliography}{41}%
\makeatletter
\providecommand \@ifxundefined [1]{%
 \@ifx{#1\undefined}
}%
\providecommand \@ifnum [1]{%
 \ifnum #1\expandafter \@firstoftwo
 \else \expandafter \@secondoftwo
 \fi
}%
\providecommand \@ifx [1]{%
 \ifx #1\expandafter \@firstoftwo
 \else \expandafter \@secondoftwo
 \fi
}%
\providecommand \natexlab [1]{#1}%
\providecommand \enquote  [1]{``#1''}%
\providecommand \bibnamefont  [1]{#1}%
\providecommand \bibfnamefont [1]{#1}%
\providecommand \citenamefont [1]{#1}%
\providecommand \href@noop [0]{\@secondoftwo}%
\providecommand \href [0]{\begingroup \@sanitize@url \@href}%
\providecommand \@href[1]{\@@startlink{#1}\@@href}%
\providecommand \@@href[1]{\endgroup#1\@@endlink}%
\providecommand \@sanitize@url [0]{\catcode `\\12\catcode `\$12\catcode
  `\&12\catcode `\#12\catcode `\^12\catcode `\_12\catcode `\%12\relax}%
\providecommand \@@startlink[1]{}%
\providecommand \@@endlink[0]{}%
\providecommand \url  [0]{\begingroup\@sanitize@url \@url }%
\providecommand \@url [1]{\endgroup\@href {#1}{\urlprefix }}%
\providecommand \urlprefix  [0]{URL }%
\providecommand \Eprint [0]{\href }%
\providecommand \doibase [0]{http://dx.doi.org/}%
\providecommand \selectlanguage [0]{\@gobble}%
\providecommand \bibinfo  [0]{\@secondoftwo}%
\providecommand \bibfield  [0]{\@secondoftwo}%
\providecommand \translation [1]{[#1]}%
\providecommand \BibitemOpen [0]{}%
\providecommand \bibitemStop [0]{}%
\providecommand \bibitemNoStop [0]{.\EOS\space}%
\providecommand \EOS [0]{\spacefactor3000\relax}%
\providecommand \BibitemShut  [1]{\csname bibitem#1\endcsname}%
\let\auto@bib@innerbib\@empty
\bibitem [{\citenamefont {Wilson}(1931)}]{Wilson1931}%
  \BibitemOpen
  \bibfield  {author} {\bibinfo {author} {\bibfnamefont {A.~H.}\ \bibnamefont
  {Wilson}},\ }\href {\doibase 10.1098/rspa.1931.0162} {\bibfield  {journal}
  {\bibinfo  {journal} {Proceedings of the Royal Society A: Mathematical,
  Physical and Engineering Sciences}\ }\textbf {\bibinfo {volume} {133}},\
  \bibinfo {pages} {458} (\bibinfo {year} {1931})}\BibitemShut {NoStop}%
\bibitem [{\citenamefont {de~Boer}\ and\ \citenamefont
  {Verwey}(1937)}]{Boer1937}%
  \BibitemOpen
  \bibfield  {author} {\bibinfo {author} {\bibfnamefont {J.~H.}\ \bibnamefont
  {de~Boer}}\ and\ \bibinfo {author} {\bibfnamefont {E.~J.~W.}\ \bibnamefont
  {Verwey}},\ }\href {\doibase 10.1088/0959-5309/49/4S/307} {\bibfield
  {journal} {\bibinfo  {journal} {Proceedings of the Physical Society}\
  }\textbf {\bibinfo {volume} {49}},\ \bibinfo {pages} {59} (\bibinfo {year}
  {1937})}\BibitemShut {NoStop}%
\bibitem [{\citenamefont {Mott}(1968)}]{Mott1968}%
  \BibitemOpen
  \bibfield  {author} {\bibinfo {author} {\bibfnamefont {N.~F.}\ \bibnamefont
  {Mott}},\ }\href {\doibase 10.1103/RevModPhys.40.677} {\bibfield  {journal}
  {\bibinfo  {journal} {Reviews of Modern Physics}\ }\textbf {\bibinfo {volume}
  {40}},\ \bibinfo {pages} {677} (\bibinfo {year} {1968})}\BibitemShut
  {NoStop}%
\bibitem [{\citenamefont {Imada}\ \emph {et~al.}(1998)\citenamefont {Imada},
  \citenamefont {Fujimori},\ and\ \citenamefont {Tokura}}]{Imada1998}%
  \BibitemOpen
  \bibfield  {author} {\bibinfo {author} {\bibfnamefont {M.}~\bibnamefont
  {Imada}}, \bibinfo {author} {\bibfnamefont {A.}~\bibnamefont {Fujimori}}, \
  and\ \bibinfo {author} {\bibfnamefont {Y.}~\bibnamefont {Tokura}},\ }\href
  {\doibase 10.1103/RevModPhys.70.1039} {\bibfield  {journal} {\bibinfo
  {journal} {Reviews of Modern Physics}\ }\textbf {\bibinfo {volume} {70}},\
  \bibinfo {pages} {1039} (\bibinfo {year} {1998})}\BibitemShut {NoStop}%
\bibitem [{\citenamefont {Lee}\ \emph {et~al.}(2008)\citenamefont {Lee},
  \citenamefont {Fursina}, \citenamefont {Mayo}, \citenamefont {Yavuz},
  \citenamefont {Colvin}, \citenamefont {Sofin}, \citenamefont {Shvets},\ and\
  \citenamefont {Natelson}}]{Lee2008}%
  \BibitemOpen
  \bibfield  {author} {\bibinfo {author} {\bibfnamefont {S.}~\bibnamefont
  {Lee}}, \bibinfo {author} {\bibfnamefont {A.}~\bibnamefont {Fursina}},
  \bibinfo {author} {\bibfnamefont {J.~T.}\ \bibnamefont {Mayo}}, \bibinfo
  {author} {\bibfnamefont {C.~T.}\ \bibnamefont {Yavuz}}, \bibinfo {author}
  {\bibfnamefont {V.~L.}\ \bibnamefont {Colvin}}, \bibinfo {author}
  {\bibfnamefont {R.~G.~S.}\ \bibnamefont {Sofin}}, \bibinfo {author}
  {\bibfnamefont {I.~V.}\ \bibnamefont {Shvets}}, \ and\ \bibinfo {author}
  {\bibfnamefont {D.}~\bibnamefont {Natelson}},\ }\href {\doibase
  10.1038/nmat2084} {\bibfield  {journal} {\bibinfo  {journal} {Nature
  materials}\ }\textbf {\bibinfo {volume} {7}},\ \bibinfo {pages} {130}
  (\bibinfo {year} {2008})}\BibitemShut {NoStop}%
\bibitem [{\citenamefont {Fursina}\ \emph {et~al.}(2009)\citenamefont
  {Fursina}, \citenamefont {Sofin}, \citenamefont {Shvets},\ and\ \citenamefont
  {Natelson}}]{Fursina2009}%
  \BibitemOpen
  \bibfield  {author} {\bibinfo {author} {\bibfnamefont {A.}~\bibnamefont
  {Fursina}}, \bibinfo {author} {\bibfnamefont {R.~G.~S.}\ \bibnamefont
  {Sofin}}, \bibinfo {author} {\bibfnamefont {I.~V.}\ \bibnamefont {Shvets}}, \
  and\ \bibinfo {author} {\bibfnamefont {D.}~\bibnamefont {Natelson}},\ }\href
  {\doibase 10.1103/PhysRevB.79.245131} {\bibfield  {journal} {\bibinfo
  {journal} {Physical Review B}\ }\textbf {\bibinfo {volume} {79}},\ \bibinfo
  {pages} {245131} (\bibinfo {year} {2009})}\BibitemShut {NoStop}%
\bibitem [{\citenamefont {Fursina}\ \emph
  {et~al.}(2010{\natexlab{a}})\citenamefont {Fursina}, \citenamefont {Sofin},
  \citenamefont {Shvets},\ and\ \citenamefont {Natelson}}]{Fursina2010}%
  \BibitemOpen
  \bibfield  {author} {\bibinfo {author} {\bibfnamefont {A.}~\bibnamefont
  {Fursina}}, \bibinfo {author} {\bibfnamefont {R.~G.~S.}\ \bibnamefont
  {Sofin}}, \bibinfo {author} {\bibfnamefont {I.~V.}\ \bibnamefont {Shvets}}, \
  and\ \bibinfo {author} {\bibfnamefont {D.}~\bibnamefont {Natelson}},\ }\href
  {\doibase 10.1103/PhysRevB.82.245112} {\bibfield  {journal} {\bibinfo
  {journal} {Physical Review B}\ }\textbf {\bibinfo {volume} {82}},\ \bibinfo
  {pages} {245112} (\bibinfo {year} {2010}{\natexlab{a}})}\BibitemShut
  {NoStop}%
\bibitem [{\citenamefont {Fursina}\ \emph
  {et~al.}(2010{\natexlab{b}})\citenamefont {Fursina}, \citenamefont {Sofin},
  \citenamefont {Shvets},\ and\ \citenamefont {Natelson}}]{Fursina2010a}%
  \BibitemOpen
  \bibfield  {author} {\bibinfo {author} {\bibfnamefont {A.}~\bibnamefont
  {Fursina}}, \bibinfo {author} {\bibfnamefont {R.~G.~S.}\ \bibnamefont
  {Sofin}}, \bibinfo {author} {\bibfnamefont {I.~V.}\ \bibnamefont {Shvets}}, \
  and\ \bibinfo {author} {\bibfnamefont {D.}~\bibnamefont {Natelson}},\ }\href
  {\doibase 10.1103/PhysRevB.81.045123} {\bibfield  {journal} {\bibinfo
  {journal} {Physical Review B}\ }\textbf {\bibinfo {volume} {81}},\ \bibinfo
  {pages} {045123} (\bibinfo {year} {2010}{\natexlab{b}})}\BibitemShut
  {NoStop}%
\bibitem [{\citenamefont {Fursina}\ \emph {et~al.}(2012)\citenamefont
  {Fursina}, \citenamefont {Sofin}, \citenamefont {Shvets},\ and\ \citenamefont
  {Natelson}}]{Fursina2012}%
  \BibitemOpen
  \bibfield  {author} {\bibinfo {author} {\bibfnamefont {A.}~\bibnamefont
  {Fursina}}, \bibinfo {author} {\bibfnamefont {R.~G.~S.}\ \bibnamefont
  {Sofin}}, \bibinfo {author} {\bibfnamefont {I.~V.}\ \bibnamefont {Shvets}}, \
  and\ \bibinfo {author} {\bibfnamefont {D.}~\bibnamefont {Natelson}},\ }\href
  {\doibase 10.1088/1367-2630/14/1/013019} {\bibfield  {journal} {\bibinfo
  {journal} {New Journal of Physics}\ }\textbf {\bibinfo {volume} {14}},\
  \bibinfo {pages} {013019} (\bibinfo {year} {2012})}\BibitemShut {NoStop}%
\bibitem [{\citenamefont {Edwards}\ and\ \citenamefont
  {Rao}(1985)}]{Edwards1985}%
  \BibitemOpen
  \bibfield  {author} {\bibinfo {author} {\bibfnamefont {P.~P.}\ \bibnamefont
  {Edwards}}\ and\ \bibinfo {author} {\bibfnamefont {C.~N.~R.}\ \bibnamefont
  {Rao}},\ }\href {http://books.google.com/books?id=zpdTAAAAMAAJ\&pgis=1}
  {\emph {\bibinfo {title} {{The Metallic and nonmetallic states of matter}}}}\
  (\bibinfo  {publisher} {Taylor \& Francis},\ \bibinfo {year} {1985})\ p.\
  \bibinfo {pages} {427}\BibitemShut {NoStop}%
\bibitem [{\citenamefont {Garc\'{\i}a}\ and\ \citenamefont
  {Sub\'{\i}as}(2004)}]{Garcia2004}%
  \BibitemOpen
  \bibfield  {author} {\bibinfo {author} {\bibfnamefont {J.}~\bibnamefont
  {Garc\'{\i}a}}\ and\ \bibinfo {author} {\bibfnamefont {G.}~\bibnamefont
  {Sub\'{\i}as}},\ }\href {\doibase 10.1088/0953-8984/16/7/R01} {\bibfield
  {journal} {\bibinfo  {journal} {Journal of Physics: Condensed Matter}\
  }\textbf {\bibinfo {volume} {16}},\ \bibinfo {pages} {R145} (\bibinfo {year}
  {2004})}\BibitemShut {NoStop}%
\bibitem [{\citenamefont {Walz}(2002)}]{Walz2002}%
  \BibitemOpen
  \bibfield  {author} {\bibinfo {author} {\bibfnamefont {F.}~\bibnamefont
  {Walz}},\ }\href {\doibase 10.1088/0953-8984/14/12/203} {\bibfield  {journal}
  {\bibinfo  {journal} {Journal of Physics: Condensed Matter}\ }\textbf
  {\bibinfo {volume} {14}},\ \bibinfo {pages} {R285} (\bibinfo {year}
  {2002})}\BibitemShut {NoStop}%
\bibitem [{\citenamefont {Kuipers}\ and\ \citenamefont
  {Brabers}(1979)}]{Kuipers1979}%
  \BibitemOpen
  \bibfield  {author} {\bibinfo {author} {\bibfnamefont {A.}~\bibnamefont
  {Kuipers}}\ and\ \bibinfo {author} {\bibfnamefont {V.~A.~M.}\ \bibnamefont
  {Brabers}},\ }\href {\doibase 10.1103/PhysRevB.20.594} {\bibfield  {journal}
  {\bibinfo  {journal} {Physical Review B}\ }\textbf {\bibinfo {volume} {20}},\
  \bibinfo {pages} {594} (\bibinfo {year} {1979})}\BibitemShut {NoStop}%
\bibitem [{\citenamefont {Ioffe}\ and\ \citenamefont
  {Regel}(1960)}]{Ioffe1960}%
  \BibitemOpen
  \bibfield  {author} {\bibinfo {author} {\bibfnamefont {A.~F.}\ \bibnamefont
  {Ioffe}}\ and\ \bibinfo {author} {\bibfnamefont {A.~R.}\ \bibnamefont
  {Regel}},\ }\href@noop {} {\bibfield  {journal} {\bibinfo  {journal} {Prog.
  Semicond.}\ }\textbf {\bibinfo {volume} {4}},\ \bibinfo {pages} {237}
  (\bibinfo {year} {1960})}\BibitemShut {NoStop}%
\bibitem [{\citenamefont {Verwey}\ and\ \citenamefont
  {Heilmann}(1947)}]{Verwey1947}%
  \BibitemOpen
  \bibfield  {author} {\bibinfo {author} {\bibfnamefont {E.~J.~W.}\
  \bibnamefont {Verwey}}\ and\ \bibinfo {author} {\bibfnamefont {E.~L.}\
  \bibnamefont {Heilmann}},\ }\href {\doibase 10.1063/1.1746464} {\bibfield
  {journal} {\bibinfo  {journal} {The Journal of Chemical Physics}\ }\textbf
  {\bibinfo {volume} {15}},\ \bibinfo {pages} {174} (\bibinfo {year}
  {1947})}\BibitemShut {NoStop}%
\bibitem [{\citenamefont {Verwey}(1939)}]{VERWEY1939}%
  \BibitemOpen
  \bibfield  {author} {\bibinfo {author} {\bibfnamefont {E.~J.~W.}\
  \bibnamefont {Verwey}},\ }\href {\doibase 10.1038/144327b0} {\bibfield
  {journal} {\bibinfo  {journal} {Nature}\ }\textbf {\bibinfo {volume} {144}},\
  \bibinfo {pages} {327} (\bibinfo {year} {1939})}\BibitemShut {NoStop}%
\bibitem [{\citenamefont {Cullen}(1970)}]{Cullen1970}%
  \BibitemOpen
  \bibfield  {author} {\bibinfo {author} {\bibfnamefont {J.~R.}\ \bibnamefont
  {Cullen}},\ }\href {\doibase 10.1063/1.1658998} {\bibfield  {journal}
  {\bibinfo  {journal} {Journal of Applied Physics}\ }\textbf {\bibinfo
  {volume} {41}},\ \bibinfo {pages} {879} (\bibinfo {year} {1970})}\BibitemShut
  {NoStop}%
\bibitem [{\citenamefont {Mott}(1980)}]{Mott1980}%
  \BibitemOpen
  \bibfield  {author} {\bibinfo {author} {\bibfnamefont {N.~F.}\ \bibnamefont
  {Mott}},\ }\href {\doibase 10.1080/01418638008221874} {\bibfield  {journal}
  {\bibinfo  {journal} {Philosophical Magazine Part B}\ }\textbf {\bibinfo
  {volume} {42}},\ \bibinfo {pages} {327} (\bibinfo {year} {1980})}\BibitemShut
  {NoStop}%
\bibitem [{\citenamefont {Austin}\ and\ \citenamefont
  {Mott}(1969)}]{Austin1969}%
  \BibitemOpen
  \bibfield  {author} {\bibinfo {author} {\bibfnamefont {I.~G.}\ \bibnamefont
  {Austin}}\ and\ \bibinfo {author} {\bibfnamefont {N.~F.}\ \bibnamefont
  {Mott}},\ }\href {\doibase 10.1080/00018736900101267} {\bibfield  {journal}
  {\bibinfo  {journal} {Advances in Physics}\ }\textbf {\bibinfo {volume}
  {18}},\ \bibinfo {pages} {41} (\bibinfo {year} {1969})}\BibitemShut {NoStop}%
\bibitem [{\citenamefont {Chakraverty}(1974)}]{Chakraverty1974}%
  \BibitemOpen
  \bibfield  {author} {\bibinfo {author} {\bibfnamefont {B.}~\bibnamefont
  {Chakraverty}},\ }\href {\doibase 10.1016/0038-1098(74)91360-X} {\bibfield
  {journal} {\bibinfo  {journal} {Solid State Communications}\ }\textbf
  {\bibinfo {volume} {15}},\ \bibinfo {pages} {1271} (\bibinfo {year}
  {1974})}\BibitemShut {NoStop}%
\bibitem [{\citenamefont {Yamada}(1980)}]{Yamada1980}%
  \BibitemOpen
  \bibfield  {author} {\bibinfo {author} {\bibfnamefont {Y.}~\bibnamefont
  {Yamada}},\ }\href {\doibase 10.1080/01418638008221877} {\bibfield  {journal}
  {\bibinfo  {journal} {Philosophical Magazine Part B}\ }\textbf {\bibinfo
  {volume} {42}},\ \bibinfo {pages} {377} (\bibinfo {year} {1980})}\BibitemShut
  {NoStop}%
\bibitem [{\citenamefont {Mott}\ and\ \citenamefont
  {Zinamon}(1970)}]{Mott1970}%
  \BibitemOpen
  \bibfield  {author} {\bibinfo {author} {\bibfnamefont {N.~F.}\ \bibnamefont
  {Mott}}\ and\ \bibinfo {author} {\bibfnamefont {Z.}~\bibnamefont {Zinamon}},\
  }\href {\doibase 10.1088/0034-4885/33/3/302} {\bibfield  {journal} {\bibinfo
  {journal} {Reports on Progress in Physics}\ }\textbf {\bibinfo {volume}
  {33}},\ \bibinfo {pages} {881} (\bibinfo {year} {1970})}\BibitemShut
  {NoStop}%
\bibitem [{\citenamefont {Wigner}(1938)}]{Wigner1938}%
  \BibitemOpen
  \bibfield  {author} {\bibinfo {author} {\bibfnamefont {E.}~\bibnamefont
  {Wigner}},\ }\href {\doibase 10.1039/tf9383400678} {\bibfield  {journal}
  {\bibinfo  {journal} {Transactions of the Faraday Society}\ }\textbf
  {\bibinfo {volume} {34}},\ \bibinfo {pages} {678} (\bibinfo {year}
  {1938})}\BibitemShut {NoStop}%
\bibitem [{\citenamefont {Park}\ \emph {et~al.}(1998)\citenamefont {Park},
  \citenamefont {Ishikawa},\ and\ \citenamefont {Tokura}}]{Park1998}%
  \BibitemOpen
  \bibfield  {author} {\bibinfo {author} {\bibfnamefont {S.}~\bibnamefont
  {Park}}, \bibinfo {author} {\bibfnamefont {T.}~\bibnamefont {Ishikawa}}, \
  and\ \bibinfo {author} {\bibfnamefont {Y.}~\bibnamefont {Tokura}},\ }\href
  {\doibase 10.1103/PhysRevB.58.3717} {\bibfield  {journal} {\bibinfo
  {journal} {Physical Review B}\ }\textbf {\bibinfo {volume} {58}},\ \bibinfo
  {pages} {3717} (\bibinfo {year} {1998})}\BibitemShut {NoStop}%
\bibitem [{\citenamefont {Gasparov}\ \emph {et~al.}(2000)\citenamefont
  {Gasparov}, \citenamefont {Tanner}, \citenamefont {Romero}, \citenamefont
  {Berger}, \citenamefont {Margaritondo},\ and\ \citenamefont
  {Forr\'{o}}}]{Gasparov2000}%
  \BibitemOpen
  \bibfield  {author} {\bibinfo {author} {\bibfnamefont {L.}~\bibnamefont
  {Gasparov}}, \bibinfo {author} {\bibfnamefont {D.}~\bibnamefont {Tanner}},
  \bibinfo {author} {\bibfnamefont {D.}~\bibnamefont {Romero}}, \bibinfo
  {author} {\bibfnamefont {H.}~\bibnamefont {Berger}}, \bibinfo {author}
  {\bibfnamefont {G.}~\bibnamefont {Margaritondo}}, \ and\ \bibinfo {author}
  {\bibfnamefont {L.}~\bibnamefont {Forr\'{o}}},\ }\href {\doibase
  10.1103/PhysRevB.62.7939} {\bibfield  {journal} {\bibinfo  {journal}
  {Physical Review B}\ }\textbf {\bibinfo {volume} {62}},\ \bibinfo {pages}
  {7939} (\bibinfo {year} {2000})}\BibitemShut {NoStop}%
\bibitem [{\citenamefont {Pimenov}\ \emph {et~al.}(2005)\citenamefont
  {Pimenov}, \citenamefont {Tachos}, \citenamefont {Rudolf}, \citenamefont
  {Loidl}, \citenamefont {Schrupp}, \citenamefont {Sing}, \citenamefont
  {Claessen},\ and\ \citenamefont {Brabers}}]{Pimenov2005}%
  \BibitemOpen
  \bibfield  {author} {\bibinfo {author} {\bibfnamefont {A.}~\bibnamefont
  {Pimenov}}, \bibinfo {author} {\bibfnamefont {S.}~\bibnamefont {Tachos}},
  \bibinfo {author} {\bibfnamefont {T.}~\bibnamefont {Rudolf}}, \bibinfo
  {author} {\bibfnamefont {A.}~\bibnamefont {Loidl}}, \bibinfo {author}
  {\bibfnamefont {D.}~\bibnamefont {Schrupp}}, \bibinfo {author} {\bibfnamefont
  {M.}~\bibnamefont {Sing}}, \bibinfo {author} {\bibfnamefont {R.}~\bibnamefont
  {Claessen}}, \ and\ \bibinfo {author} {\bibfnamefont {V.~A.~M.}\ \bibnamefont
  {Brabers}},\ }\href {\doibase 10.1103/PhysRevB.72.035131} {\bibfield
  {journal} {\bibinfo  {journal} {Physical Review B}\ }\textbf {\bibinfo
  {volume} {72}},\ \bibinfo {pages} {035131} (\bibinfo {year}
  {2005})}\BibitemShut {NoStop}%
\bibitem [{\citenamefont {Devreese}\ and\ \citenamefont
  {Alexandrov}(2009)}]{Devreese2009}%
  \BibitemOpen
  \bibfield  {author} {\bibinfo {author} {\bibfnamefont {J.~T.}\ \bibnamefont
  {Devreese}}\ and\ \bibinfo {author} {\bibfnamefont {A.~S.}\ \bibnamefont
  {Alexandrov}},\ }\href {\doibase 10.1088/0034-4885/72/6/066501} {\bibfield
  {journal} {\bibinfo  {journal} {Reports on Progress in Physics}\ }\textbf
  {\bibinfo {volume} {72}},\ \bibinfo {pages} {066501} (\bibinfo {year}
  {2009})}\BibitemShut {NoStop}%
\bibitem [{\citenamefont {Basov}\ \emph {et~al.}(2011)\citenamefont {Basov},
  \citenamefont {Averitt}, \citenamefont {van~der Marel}, \citenamefont
  {Dressel},\ and\ \citenamefont {Haule}}]{Basov2011}%
  \BibitemOpen
  \bibfield  {author} {\bibinfo {author} {\bibfnamefont {D.~N.}\ \bibnamefont
  {Basov}}, \bibinfo {author} {\bibfnamefont {R.~D.}\ \bibnamefont {Averitt}},
  \bibinfo {author} {\bibfnamefont {D.}~\bibnamefont {van~der Marel}}, \bibinfo
  {author} {\bibfnamefont {M.}~\bibnamefont {Dressel}}, \ and\ \bibinfo
  {author} {\bibfnamefont {K.}~\bibnamefont {Haule}},\ }\href {\doibase
  10.1103/RevModPhys.83.471} {\bibfield  {journal} {\bibinfo  {journal}
  {Reviews of Modern Physics}\ }\textbf {\bibinfo {volume} {83}},\ \bibinfo
  {pages} {471} (\bibinfo {year} {2011})}\BibitemShut {NoStop}%
\bibitem [{\citenamefont {Park}\ \emph {et~al.}(1997)\citenamefont {Park},
  \citenamefont {Allen}, \citenamefont {Metcalf},\ and\ \citenamefont
  {Chen}}]{Park1997}%
  \BibitemOpen
  \bibfield  {author} {\bibinfo {author} {\bibfnamefont {J.-H.}\ \bibnamefont
  {Park}}, \bibinfo {author} {\bibfnamefont {J.~W.}\ \bibnamefont {Allen}},
  \bibinfo {author} {\bibfnamefont {P.}~\bibnamefont {Metcalf}}, \ and\
  \bibinfo {author} {\bibfnamefont {C.~T.}\ \bibnamefont {Chen}},\ }\href
  {\doibase 10.1103/PhysRevB.55.12813} {\bibfield  {journal} {\bibinfo
  {journal} {Physical Review B}\ }\textbf {\bibinfo {volume} {55}},\ \bibinfo
  {pages} {12813} (\bibinfo {year} {1997})}\BibitemShut {NoStop}%
\bibitem [{\citenamefont {Schrupp}\ \emph {et~al.}(2005)\citenamefont
  {Schrupp}, \citenamefont {Sing}, \citenamefont {Tsunekawa}, \citenamefont
  {Fujiwara}, \citenamefont {Kasai}, \citenamefont {Sekiyama}, \citenamefont
  {Suga}, \citenamefont {Muro}, \citenamefont {Brabers},\ and\ \citenamefont
  {Claessen}}]{Schrupp2005}%
  \BibitemOpen
  \bibfield  {author} {\bibinfo {author} {\bibfnamefont {D.}~\bibnamefont
  {Schrupp}}, \bibinfo {author} {\bibfnamefont {M.}~\bibnamefont {Sing}},
  \bibinfo {author} {\bibfnamefont {M.}~\bibnamefont {Tsunekawa}}, \bibinfo
  {author} {\bibfnamefont {H.}~\bibnamefont {Fujiwara}}, \bibinfo {author}
  {\bibfnamefont {S.}~\bibnamefont {Kasai}}, \bibinfo {author} {\bibfnamefont
  {A.}~\bibnamefont {Sekiyama}}, \bibinfo {author} {\bibfnamefont
  {S.}~\bibnamefont {Suga}}, \bibinfo {author} {\bibfnamefont {T.}~\bibnamefont
  {Muro}}, \bibinfo {author} {\bibfnamefont {V.~A.~M.}\ \bibnamefont
  {Brabers}}, \ and\ \bibinfo {author} {\bibfnamefont {R.}~\bibnamefont
  {Claessen}},\ }\href {\doibase 10.1209/epl/i2005-10045-y} {\bibfield
  {journal} {\bibinfo  {journal} {Europhysics Letters (EPL)}\ }\textbf
  {\bibinfo {volume} {70}},\ \bibinfo {pages} {789} (\bibinfo {year}
  {2005})}\BibitemShut {NoStop}%
\bibitem [{\citenamefont {Poddar}\ \emph {et~al.}(2003)\citenamefont {Poddar},
  \citenamefont {Fried}, \citenamefont {Markovich}, \citenamefont {Sharoni},
  \citenamefont {Katz}, \citenamefont {Wizansky},\ and\ \citenamefont
  {Millo}}]{Poddar2003}%
  \BibitemOpen
  \bibfield  {author} {\bibinfo {author} {\bibfnamefont {P.}~\bibnamefont
  {Poddar}}, \bibinfo {author} {\bibfnamefont {T.}~\bibnamefont {Fried}},
  \bibinfo {author} {\bibfnamefont {G.}~\bibnamefont {Markovich}}, \bibinfo
  {author} {\bibfnamefont {A.}~\bibnamefont {Sharoni}}, \bibinfo {author}
  {\bibfnamefont {D.}~\bibnamefont {Katz}}, \bibinfo {author} {\bibfnamefont
  {T.}~\bibnamefont {Wizansky}}, \ and\ \bibinfo {author} {\bibfnamefont
  {O.}~\bibnamefont {Millo}},\ }\href {\doibase 10.1209/epl/i2003-00141-0}
  {\bibfield  {journal} {\bibinfo  {journal} {Europhysics Letters (EPL)}\
  }\textbf {\bibinfo {volume} {64}},\ \bibinfo {pages} {98} (\bibinfo {year}
  {2003})}\BibitemShut {NoStop}%
\bibitem [{\citenamefont {Jordan}\ \emph {et~al.}(2006)\citenamefont {Jordan},
  \citenamefont {Cazacu}, \citenamefont {Manai}, \citenamefont {Ceballos},
  \citenamefont {Murphy},\ and\ \citenamefont {Shvets}}]{Jordan2006}%
  \BibitemOpen
  \bibfield  {author} {\bibinfo {author} {\bibfnamefont {K.}~\bibnamefont
  {Jordan}}, \bibinfo {author} {\bibfnamefont {A.}~\bibnamefont {Cazacu}},
  \bibinfo {author} {\bibfnamefont {G.}~\bibnamefont {Manai}}, \bibinfo
  {author} {\bibfnamefont {S.~F.}\ \bibnamefont {Ceballos}}, \bibinfo {author}
  {\bibfnamefont {S.}~\bibnamefont {Murphy}}, \ and\ \bibinfo {author}
  {\bibfnamefont {I.~V.}\ \bibnamefont {Shvets}},\ }\href {\doibase
  10.1103/PhysRevB.74.085416} {\bibfield  {journal} {\bibinfo  {journal}
  {Physical Review B}\ }\textbf {\bibinfo {volume} {74}},\ \bibinfo {pages}
  {085416} (\bibinfo {year} {2006})}\BibitemShut {NoStop}%
\bibitem [{\citenamefont {Chainani}\ \emph {et~al.}(1995)\citenamefont
  {Chainani}, \citenamefont {Yokoya}, \citenamefont {Morimoto}, \citenamefont
  {Takahashi},\ and\ \citenamefont {Todo}}]{Chainani1995}%
  \BibitemOpen
  \bibfield  {author} {\bibinfo {author} {\bibfnamefont {A.}~\bibnamefont
  {Chainani}}, \bibinfo {author} {\bibfnamefont {T.}~\bibnamefont {Yokoya}},
  \bibinfo {author} {\bibfnamefont {T.}~\bibnamefont {Morimoto}}, \bibinfo
  {author} {\bibfnamefont {T.}~\bibnamefont {Takahashi}}, \ and\ \bibinfo
  {author} {\bibfnamefont {S.}~\bibnamefont {Todo}},\ }\href {\doibase
  10.1103/PhysRevB.51.17976} {\bibfield  {journal} {\bibinfo  {journal}
  {Physical Review B}\ }\textbf {\bibinfo {volume} {51}},\ \bibinfo {pages}
  {17976} (\bibinfo {year} {1995})}\BibitemShut {NoStop}%
\bibitem [{\citenamefont {Mottaghizadeh}\ \emph {et~al.}(2014)\citenamefont
  {Mottaghizadeh}, \citenamefont {Yu}, \citenamefont {Lang}, \citenamefont
  {Zimmers},\ and\ \citenamefont {Aubin}}]{Mottaghizadeh2014}%
  \BibitemOpen
  \bibfield  {author} {\bibinfo {author} {\bibfnamefont {A.}~\bibnamefont
  {Mottaghizadeh}}, \bibinfo {author} {\bibfnamefont {Q.}~\bibnamefont {Yu}},
  \bibinfo {author} {\bibfnamefont {P.}~\bibnamefont {Lang}}, \bibinfo {author}
  {\bibfnamefont {A.}~\bibnamefont {Zimmers}}, \ and\ \bibinfo {author}
  {\bibfnamefont {H.}~\bibnamefont {Aubin}},\ }\href {\doibase
  10.1103/PhysRevLett.112.066803} {\bibfield  {journal} {\bibinfo  {journal}
  {Physical Review Letters}\ }\textbf {\bibinfo {volume} {112}},\ \bibinfo
  {pages} {066803} (\bibinfo {year} {2014})}\BibitemShut {NoStop}%
\bibitem [{\citenamefont {Kuemmeth}\ \emph {et~al.}(2008)\citenamefont
  {Kuemmeth}, \citenamefont {Bolotin}, \citenamefont {Shi},\ and\ \citenamefont
  {Ralph}}]{Kuemmeth2008}%
  \BibitemOpen
  \bibfield  {author} {\bibinfo {author} {\bibfnamefont {F.}~\bibnamefont
  {Kuemmeth}}, \bibinfo {author} {\bibfnamefont {K.~I.}\ \bibnamefont
  {Bolotin}}, \bibinfo {author} {\bibfnamefont {S.-F.}\ \bibnamefont {Shi}}, \
  and\ \bibinfo {author} {\bibfnamefont {D.~C.}\ \bibnamefont {Ralph}},\ }\href
  {\doibase 10.1021/nl802473n} {\bibfield  {journal} {\bibinfo  {journal} {Nano
  letters}\ }\textbf {\bibinfo {volume} {8}},\ \bibinfo {pages} {4506}
  (\bibinfo {year} {2008})}\BibitemShut {NoStop}%
\bibitem [{\citenamefont {Aubin}\ \emph {et~al.}(2002)\citenamefont {Aubin},
  \citenamefont {Greene}, \citenamefont {Jian},\ and\ \citenamefont
  {Hinks}}]{Aubin2002}%
  \BibitemOpen
  \bibfield  {author} {\bibinfo {author} {\bibfnamefont {H.}~\bibnamefont
  {Aubin}}, \bibinfo {author} {\bibfnamefont {L.~H.}\ \bibnamefont {Greene}},
  \bibinfo {author} {\bibfnamefont {S.}~\bibnamefont {Jian}}, \ and\ \bibinfo
  {author} {\bibfnamefont {D.}~\bibnamefont {Hinks}},\ }\href {\doibase
  10.1103/PhysRevLett.89.177001} {\bibfield  {journal} {\bibinfo  {journal}
  {Physical Review Letters}\ }\textbf {\bibinfo {volume} {89}},\ \bibinfo
  {pages} {177001} (\bibinfo {year} {2002})}\BibitemShut {NoStop}%
\bibitem [{\citenamefont {Vincent}\ \emph {et~al.}(2012)\citenamefont
  {Vincent}, \citenamefont {Klyatskaya}, \citenamefont {Ruben}, \citenamefont
  {Wernsdorfer},\ and\ \citenamefont {Balestro}}]{Vincent2012}%
  \BibitemOpen
  \bibfield  {author} {\bibinfo {author} {\bibfnamefont {R.}~\bibnamefont
  {Vincent}}, \bibinfo {author} {\bibfnamefont {S.}~\bibnamefont {Klyatskaya}},
  \bibinfo {author} {\bibfnamefont {M.}~\bibnamefont {Ruben}}, \bibinfo
  {author} {\bibfnamefont {W.}~\bibnamefont {Wernsdorfer}}, \ and\ \bibinfo
  {author} {\bibfnamefont {F.}~\bibnamefont {Balestro}},\ }\href {\doibase
  10.1038/nature11341} {\bibfield  {journal} {\bibinfo  {journal} {Nature}\
  }\textbf {\bibinfo {volume} {488}},\ \bibinfo {pages} {357} (\bibinfo {year}
  {2012})}\BibitemShut {NoStop}%
\bibitem [{Sup()}]{SupplementalMaterial}%
  \BibitemOpen
  \href@noop {} {\bibinfo  {journal} {See Supplemental Material at [URL] for
  details on materials used for synthesis and characterization techniques}\
  }\BibitemShut {NoStop}%
\bibitem [{\citenamefont {Yu}\ \emph {et~al.}(2013)\citenamefont {Yu},
  \citenamefont {Cui}, \citenamefont {Lequeux}, \citenamefont {Zimmers},
  \citenamefont {Ulysse}, \citenamefont {Rebuttini}, \citenamefont {Pinna},\
  and\ \citenamefont {Aubin}}]{Yu2013b}%
  \BibitemOpen
\bibfield  {journal} {  }\bibfield  {author} {\bibinfo {author} {\bibfnamefont
  {Q.}~\bibnamefont {Yu}}, \bibinfo {author} {\bibfnamefont {L.}~\bibnamefont
  {Cui}}, \bibinfo {author} {\bibfnamefont {N.}~\bibnamefont {Lequeux}},
  \bibinfo {author} {\bibfnamefont {A.}~\bibnamefont {Zimmers}}, \bibinfo
  {author} {\bibfnamefont {C.}~\bibnamefont {Ulysse}}, \bibinfo {author}
  {\bibfnamefont {V.}~\bibnamefont {Rebuttini}}, \bibinfo {author}
  {\bibfnamefont {N.}~\bibnamefont {Pinna}}, \ and\ \bibinfo {author}
  {\bibfnamefont {H.}~\bibnamefont {Aubin}},\ }\href {\doibase
  10.1021/nn305264g} {\bibfield  {journal} {\bibinfo  {journal} {ACS nano}\
  }\textbf {\bibinfo {volume} {7}},\ \bibinfo {pages} {1487} (\bibinfo {year}
  {2013})}\BibitemShut {NoStop}%
\bibitem [{\citenamefont {Beebe}\ \emph {et~al.}(2006)\citenamefont {Beebe},
  \citenamefont {Kim}, \citenamefont {Gadzuk}, \citenamefont {Frisbie},
  \citenamefont {Kushmerick},\ and\ \citenamefont {{Daniel
  Frisbie}}}]{Beebe2006}%
  \BibitemOpen
  \bibfield  {author} {\bibinfo {author} {\bibfnamefont {J.~M.}\ \bibnamefont
  {Beebe}}, \bibinfo {author} {\bibfnamefont {B.}~\bibnamefont {Kim}}, \bibinfo
  {author} {\bibfnamefont {J.~W.}\ \bibnamefont {Gadzuk}}, \bibinfo {author}
  {\bibfnamefont {C.~D.}\ \bibnamefont {Frisbie}}, \bibinfo {author}
  {\bibfnamefont {J.~G.}\ \bibnamefont {Kushmerick}}, \ and\ \bibinfo {author}
  {\bibfnamefont {C.}~\bibnamefont {{Daniel Frisbie}}},\ }\href {\doibase
  10.1103/PhysRevLett.97.026801} {\bibfield  {journal} {\bibinfo  {journal}
  {Physical Review Letters}\ }\textbf {\bibinfo {volume} {97}},\ \bibinfo
  {pages} {1} (\bibinfo {year} {2006})}\BibitemShut {NoStop}%
\bibitem [{\citenamefont {Lundin}\ and\ \citenamefont
  {McKenzie}(2002)}]{Lundin2002}%
  \BibitemOpen
  \bibfield  {author} {\bibinfo {author} {\bibfnamefont {U.}~\bibnamefont
  {Lundin}}\ and\ \bibinfo {author} {\bibfnamefont {R.}~\bibnamefont
  {McKenzie}},\ }\href {\doibase 10.1103/PhysRevB.66.075303} {\bibfield
  {journal} {\bibinfo  {journal} {Physical Review B}\ }\textbf {\bibinfo
  {volume} {66}},\ \bibinfo {pages} {075303} (\bibinfo {year}
  {2002})}\BibitemShut {NoStop}%
\end{thebibliography}%

\end{document}